%% file: paper.tex
\documentstyle[multicol,aps,prb]{revtex}
\input{uzspecial}

\newcommand{\ket}[1]{\left | \, #1 \right \rangle}
\newcommand{\bra}[1]{\left \langle #1 \, \right |}

\newcommand{\beq}{\begin{equation}}
\newcommand{\eeq}{\end{equation}}

\begin{document}

\draft

\preprint{IUCM98-001}

\title{Periphery deformations and tunneling at correlated
quantum-Hall edges}

\author{U.~Z\"ulicke\cite{preadd} and A.~H.~MacDonald}

\address{Department of Physics, Indiana University, Bloomington,
Indiana 47405}

\date{\today}

\maketitle

\begin{abstract}

We argue that, at any filling factor $\nu$, correlated
quantum-Hall systems possess a set of chiral boson excitations
which are generated by electronically rigid deformations of the
system's periphery. We submit that tunneling electrons can be
accommodated, at low energies, in these systems only by
periphery-deformation excitations. This property would explain
the recent observation of a tunneling density of states at the
edge which does not exhibit a strong dependence on the occurrence
or absence of the quantum Hall effect and has a power-law 
dependence on energy with exponent $\nu^{-1}-1$.
 
\end{abstract}


\narrowtext

Experimental and theoretical studies of two-dimensional (2D) 
electron systems in strong perpendicular magnetic fields have 
uncovered a rich body of many-particle physics.\cite{ahm:ssc:97}
For magnetic fields sufficiently strong and densities
sufficiently low, electrons occupy only states in the lowest
Landau level (LLL) which have identical kinetic
energies.\cite{spinapology} In this zero-band-width correlation
problem, interactions alone control the low-energy physics.
We refer to this strong-field regime for 2D electron systems as
the quantum-Hall regime, after the transport anomaly that occurs
at certain fractional Landau-level filling factors $\nu$, 
including $\nu = 1/m$ with $m$ odd. The quantum Hall effect
occurs\cite{ahmintro} when the electronic system is
incompressible, i.e., when there is a finite energy gap for
particle-hole excitations in the bulk and
low-lying excitations are localized at the boundary of the finite
2D electron system.\cite{caveatlocalized} The edge electrons
then constitute a particularly intriguing realization of a
one-dimensional (1D) electron system.\cite{voit:reprog:94} In
the simplest case, they are described by a
{\em chiral-Luttinger-liquid\/} ($\chi$LL)\cite{wen:prb:90}
model which contains only the right-moving degrees of freedom of
a conventional 1D electron system. The chiral attribute is
permitted because the magnetic field breaks time-reversal
symmetry. The $\chi$LL model predicts that the electronic
tunneling density of states (TDOS) at fractional filling factors
vanishes as a power law at small energies. The simple $\chi$LL
model is expected to apply only for $\nu = 1/m$ with $m$ odd and,
even then, only when the 2D electron system has a sharp edge.
More generally,\cite{ahm:prl:90,wen} the edge system is expected
to be described by generalized 1D electron models with several
branches of chiral 1D excitations present.

The present work is motivated by recent experimental studies of 
tunneling into fractional-quantum-Hall (FQH) edges. At filling
factor $\nu = 1/3$, the low-energy TDOS was
found\cite{amc:prl:96} to vary approximately as 
\begin{equation}\label{powerlaw}
A^>( \varepsilon) \propto \varepsilon^{\frac{1}{\nu}-1} \quad ,
\end{equation}
in agreement with predictions based on $\chi$LL theory. However,
subsequent experiments\cite{amc:preprint:97} found that the 
TDOS is described by Eq.~(\ref{powerlaw}) equally well over the entire
range $1/4 \le \nu\le 1$. This experimental result requires fine tuning
of parameters in order to be consistent with generalized $\chi$LL
and other models\cite{mpaf:prb:95a,bih:prl:98} which have been
proposed for edges at $\nu \ne 1/m$. Most surprising, the
power-law behavior was found to be insensitive to the occurrence
or absence of the quantum Hall effect, even though the very use
of a 1D model for the microscopic electron physics springs from 
the excitation gap in the bulk of the 2D system. The simple
single-branch $\chi$LL model appears to apply in circumstances
where it had seemed clear that it could not apply. These
important experiments have spurred intense theoretical
effort\cite{bih:prl:98,vighan,vigprep,fisherleewen} aimed at resolving
this puzzle and identifying the character of the revision apparently
required of FQH edge-physics theory.

In this article, we propose an explanation for the experimental 
results which is based on two key observations. First, at filling
factors $\nu < 1$, strong and subtle correlations exist in the
bulk of a 2D electron gas, even when it is compressible.
Tunneling at low energies is extremely strongly suppressed
because states with an {\em uncorrelated} electron inserted or
removed are nearly orthogonal to the low-energy states in which
all electrons are strongly correlated.\cite{wigcryst} This
property is established by bulk 2D-2D tunneling
experiments\cite{jpe:prl:95} which manifest a wide gap
in the TDOS at all values of $\nu$. Like their edge counterparts,
bulk-TDOS results exhibit surprisingly little dependence on
the occurrence or absence of the quantum Hall effect. Second, in
any finite quantum-Hall system, there exists a series of
low-energy excitations which are generated by electronically
rigid deformations of the system periphery.  We will refer to
these as {\em periphery-deformation\/} (PD) excitations.  We
propose that in systems where the confining potential is smooth
enough that electrons remain strongly correlated at the edge,
tunneling electrons can be accommodated, at low energies,
only by periphery deformations.  In the following, we explain why
PD excitations are generic, and establish the properties
underlying the edge-TDOS observations.

Our analysis rests squarely on the nature of the restrictions placed on
electronic correlations by the requirement that all electrons
lie in the LLL. This constraint is
conveniently expressed in the symmetric gauge, for which the LLL
single-particle wave functions have the following simple form:
\begin{equation}
\phi_m(z) = \frac{1}{(2\pi\ell^2\, m!)^{1/2}} \, \left(
\frac{z}{2\ell}\right)^m \, \exp \left\{ - \frac{|z|^2}{4\ell^2}
\right\} \quad .
\label{singpart}
\end{equation}
Here, $z := x - i y$ is the complex 2D coordinate, $m$ is a 
non-negative integer, and $\ell:=\sqrt{\hbar c/|e B|}$ denotes
the magnetic length. We assume a finite FQH system that is
homogeneous throughout its bulk, which we take for the
moment to be a circular droplet of radius $R$. We
start by describing the low-energy excitations generated by
periphery deformations which can be expressed as linear
combinations of the following basis functions: 
\begin{mathletters}
\begin{equation}
\Psi_{\{m\}}[z] = Q_{\{m\}}[z] \,\, \Psi_0[z] \quad ,
\label{groundstatedeform}
\end{equation}
where $\Psi_0[z]$ is the ground-state many-particle wave
function, and
\begin{equation}
Q_{\{m\}}[z] = \sum_{P} z_{P_1}^{m_1} z_{P_2}^{m_2} \ldots
z_{P_N}^{m_N}
\label{Qdef}
\end{equation} 
\end{mathletters}
is a symmetric polynomial in the complex coordinates $z_j$ with
homogeneous\cite{circsym} degree $M = \sum_j m_j$. The sum is
over permutations $P$ of the particle indices. These polynomials
are in one-to-one correspondence with the many-body states of a
LLL Bose system. We restrict our attention to states of the 
form~(\ref{groundstatedeform}) for which $M \ll \sqrt{N}$, where
$N = \nu R^2 / (2\ell^2)$ is the total number of electrons in the
system.  Note that all these states have total angular momentum
greater than that of the ground state.

The properties we require follow from matrix elements of 
correlation-function operators between  
the states defined in Eq.~(\ref{groundstatedeform}). For example,
consider the one-particle density matrix for which 
\widetext
\top{-2.8cm}
\begin{mathletters}\label{opdensity}
\begin{equation}
\bra{\Psi_{\{m^\prime\}}} \rho(z,z^\prime )
\ket{\Psi_{\{m\}}} = 
\frac{N\int \prod_{j=2}^{N} d^2 z_j \,\, Q_{\{m^\prime\}}(\bar
z^\prime, \bar z_2, \ldots , \bar z_N ) \, \Psi_{0} (\bar
z^\prime, \bar z_2, \ldots , \bar z_N ) \,\, Q_{\{m\}}(z, z_2,
\ldots , z_N ) \, \Psi_0 (z,  z_2, \ldots ,  z_N )}{\left[\int
\prod_{j=1}^{N} d^2 z_j \,\, \left| \Psi_{\{m^\prime\}}[z]
\right|^2\right]^{\frac{1}{2}} \, \left[\int \prod_{j=1}^{N}
d^2 z_j \,\, \left| \Psi_{\{m\}}[z] \right|^2\right]^\frac{1}{2}}
\, .
\label{densmat}
\end{equation} 
A key observation necessary for the evaluation of these matrix elements
is that the dependence of the symmetric polynomials on the
complex coordinates is slow compared to that of $\Psi_0[z]$.
The integrals can be evaluated by a coarse-graining procedure in
which the integration areas for each coordinate are divided into
subareas whose size is large compared to the correlation
length in the electronic ground state but small compared to the
length scale on which $Q_{\{m\}}[z]$ varies significantly. This
is possible since the correlation length is $\sim \ell$, and the
typical relative change in $Q_{\{m\}}[z]$ on this scale is $\sim
m_{\text{max}} \, \ell /R \ll 1$. Here, $m_{\text{max}} \le M$ is
the maximum power of an individual complex coordinate in
$Q_{\{m\}}[z]$. Taking the symmetric polynomials outside of the
integrals within correlation areas, neglecting correlations in 
$\Psi_0[z]$ between different correlation areas, and invoking
homogeneity in the bulk of the 2D system, we obtain
\begin{equation}
\bra{\Psi_{\{m^\prime\}}} \rho(z,z^\prime )
\ket{\Psi_{\{m\}}} =
\rho_0(z,z^\prime) \,\, \frac{\prod_{k=2}^{N} \int dk \,\,
Q_{\{m^\prime\}}(\bar z^\prime, \bar z_2, \ldots , \bar z_N) \,
Q_{\{m\}}(z, z_2, \ldots , z_N )}{\left[\prod_{k=1}^N \int dk\,\,
\left| Q_{\{m^\prime\}}(z_1 , z_2, \ldots , z_N)\right|^2
\right]^\frac{1}{2} \, \left[\prod_{k=1}^N \int dk\,\, \left|
Q_{\{m\}}(z_1, z_2, \ldots , z_N)\right|^2\right]^\frac{1}{2}}
\quad ,
\label{coarsegrain}
\end{equation}
where $\rho_0(z,z')$ is the {\em expectation value} of the 
one-particle density operator in the ground state. We have
defined $\int dk$ in the above equation as the average for the
coordinate with index $k$ over the circular area inside of which
the electron system is uniform. The weighting factor for this
integral is constant because of the uniformity of the electron
system and is {\em not} $\exp(-|z_k|^2)/(2\ell^2)$ as it would
be if we were evaluating matrix elements of the density operator
between the corresponding LLL boson states.

Since $\int dk \, {\bar z_k}^{m^\prime_k} z_k^{m_k} = 
\delta_{m^\prime_k,m_k}\, R^{2 m_k}/(m_k+1)$, the matrix elements
in Eq.~(\ref{coarsegrain}) are zero unless the sets of angular
momenta in $\{m^\prime\}$ and $\{m\}$ are identical:  
\begin{equation}
\bra{\Psi_{\{m^\prime\}}} \rho(z,z^\prime ) \ket{\Psi_{\{m\}}} =
\rho_0(z,z^\prime) \,\, \left[ \frac{\sum_{P} (\bar z^\prime z /
R^2 )^{m_{P_1}} \prod_{k>1} (m_{P_k}+1)^{-1}}{N!\prod_{l}
(m_l+1)^{-1}}\right] \,\, \delta_{\{m^\prime\},\{m\}} \quad .
\label{finalresult}
\end{equation}
\end{mathletters}
Our final result then follows from the observation that, since
$\sum_{l} m_l = M \ll \sqrt{N} \ll N$, $m_{P_1}$ in
Eq.~(\ref{finalresult}) is almost always zero, allowing the
factor in square brackets to be replaced by unity 
with at least $1/\sqrt{N}$ accuracy. For any macroscopic
sample, $\rho (z,z^\prime)$ can 
hence be replaced, in the PD subspace, by a constant equal to its
expectation value in the ground state. This conclusion holds as
long as $z$ and $z^\prime$ are in the interior of the sample. 
We discuss the edge properties of these wave functions below. A
similar argument can be used to show that, within this subspace,
the two-particle correlation function, and hence the interaction
Hamiltonian, can be replaced by a constant equal to the
interaction contribution to the ground-state energy. It follows
that the energy of PD excitations is, to a good approximation,
entirely due to changes in the interaction of electrons with the
external potential confining the electrons in the sample.
\narrowtext

To make connection with 1D electron-gas models, it is useful at
this point to switch to a Hall-bar geometry and to the Landau
gauge in which the LLL orbitals are plane waves along the edge
and are labeled by a 1D wave vector $k$:
\begin{equation}
\phi_k(x,y) = \frac{\exp\{i k x\}}{L^{1/2}}\,\,\frac{\exp{\left
\{-(y -\ell^2 k)^2/(2\ell^2)\right\}}}{(\pi^{1/2}\, \ell)^{1/2}}
\quad .
\label{landaugauge}
\end{equation} 
The second-quantized operator for the 2D electron number density 
in the Landau-gauge representation is
\widetext
\top{-2.8cm}
\beq\label{2Ddensity}
\hat n^{\text{2D}}(x, y) = \frac{1}{L} \sum_q \exp\{i q x \} \,
\exp\{-(q\ell)^2/4\} \,\sum_k \frac{\exp\{-(y - k\ell^2)^2/\ell^2
\}}{\pi^{1/2}\,\ell}\,\, c^\dagger_{k+q/2} c_{k-q/2}\quad .
\eeq
\bottom{-2.7cm}
\narrowtext
\noindent
We define a 1D edge density-operator $\hat\rho^{\text{1D}}(x)$
for the 2D electron system by integrating Eq.~(\ref{2Ddensity})
from a reference point located in the bulk, at $y=-Y < 0$, across
the edge, assumed to be near $y=0$, to infinity. Its 1D Fourier
transform has the following form:
\begin{mathletters}
\begin{equation}\label{1Ddensity}
\hat\rho^{\text{1D}}_q = \exp\{-(q\ell)^2 / 4\} \sum_k I_k \,
c^{\dagger}_{k+q/2} c_{k-q/2} \quad ,
\label{rhoq1D}
\end{equation}
where 
\begin{equation} 
I_k = \frac{1}{\pi^{1/2}\,\ell} \int_{-Y}^{\infty} d y\,\, \exp
\{ - (y - \ell^2 k)^2/ \ell^2\} \quad . 
\label{I_k}
\end{equation} 
\end{mathletters}
The function $I_k$ increases monotonically with $k$, approaching
zero for $k\ell +Y/\ell\ll-1 $ and unity for $k\ell+Y/\ell\gg 1$.
In the PD subspace, we have\cite{qell} $q\ell\ll1$, allowing us
to drop the exponential factor in Eq.~(\ref{1Ddensity}).
Subsequent calculations require the following result for the edge
density operator in the PD subspace:
\begin{mathletters}
\begin{eqnarray}\label{commutator}
[\hat\rho^{\text{1D}}_{-q^\prime}\, , \, \hat\rho^{\text{1D}}_{q}
] &=& \nu \,\, \delta_{q,q^\prime} \,
\sum_k (I^2_{k+q/2} - I^2_{k-q/2} ) \,\, , \\
&=& \nu\,\frac{q L}{2\pi}\,\delta_{q,
q^\prime} \quad .
\label{commute}
\end{eqnarray} 
\end{mathletters}
Equation~(\ref{commutator}) follows since $c^{\dagger}_{k^\prime}
c_k$ is the Landau-gauge second-quantized form for the
one-particle density-matrix operator, and can therefore, in the
PD subspace, be replaced by its ground-state expectation value,
$\bra{\Psi_0} c^{\dagger}_{k^\prime} c_k \ket{\Psi_0} = \nu\,
\delta_{k^\prime, k}$. 

We have argued above that only the external-potential part of the
electron Hamiltonian contributes non-trivially to the projection
of the Hamiltonian onto the PD subspace. Linearizing the
dispersion relation for electrons in the LLL, we obtain, up to a
constant term, for the PD-excitation sector of the Hamiltonian,
\beq\label{energy}
H_{\text{PD}} = \frac{2\pi\hbar}{L} \, \frac{v_{\text{F}}}{\nu}
\sum_{q>0} \, \hat\rho^{\text{1D}}_q \, \hat\rho^{\text{1D}}_{-q}
\quad ,
\eeq
where the velocity $v_{\text{F}} = (\ell^2/\hbar) \partial
V_{\text{ext}}/ \partial y|_0 $. We see that the operators $\hat
\rho^{\text{1D}}_q$ and $\hat\rho^{\text{1D}}_{-q}$ are
proportional to creation and annihilation operators which can be
used to generate many-particle eigenstates in the PD subspace. 

We now turn to a discussion of the periphery-deformation 
states\cite{stone:prb:90} defined by 
\beq\label{edgewave}
\ket{\Psi_q(\eta)} = \exp{\left\{ i\left[ \eta\,\hat
\rho^{\text{1D}}_q + \bar\eta\,\hat\rho^{\text{1D}}_{-q} \right]
\right\}} \,\, \ket{\Psi_0} \quad ,
\eeq
which represent {\em electronically rigid\/} periphery
deformations of the 2D ground-state density profile. That the
latter is true is seen simply by calculating the 2D
electron-density profile in such a state;
\begin{mathletters}
\beq
\delta n^{\text{2D}}_q (\vec r) = \bra{\Psi_q(\eta)} \hat
n^{\text{2D}}(\vec r) \ket{\Psi_q(\eta)} - \bra{\Psi_0} \hat
n^{\text{2D}}(\vec r) \ket{\Psi_0} \,\, ,
\eeq
with the 2D coordinate vector denoted by $\vec r = (x, y)$. A
straightforward calculation yields
\beq\label{displace}
\delta n^{\text{2D}}_q (\vec r) = n^{\text{2D}}(x, y - 2 \pi
\ell^2\,\varrho^{\text{1D}}_q(x)/\nu ) - n^{\text{2D}}(x, y)
\quad ,
\eeq
\end{mathletters}
where $n^{\text{2D}}(\vec r) = \bra{\Psi_0} \hat n^{\text{2D}}
(\vec r) \ket{\Psi_0}$ is the 2D electron-density profile in the
ground state, and $\varrho^{\text{1D}}_q(x) = \bra{\Psi_q(\eta)}
\hat \rho^{\text{1D}}(x) \ket{\Psi_q(\eta)}$ is the expectation
value of the 1D density operator. 
Note that the derivation of Eq.~(\ref{displace}) requires only
the definitions of the 2D and
1D density operators [Eqs.~(\ref{2Ddensity})
and~(\ref{1Ddensity})], and momentum conservation.

To this point, we have discussed only PD excitations of the 
ground state.  At general filling factors, a large number of 
many-particle states representing excitations, not generically
localized at the surface, with strong
correlations slightly altered from those of the ground state
will occur at low energies.  PD excitations can occur for any of
these states.  The line of argument outlined above implies that 
Eq.~(\ref{commute}) holds within each of these PD subspaces and 
that there are no matrix elements of the commutator between
subspaces; the low energy Hilbert space is thus the direct product of  
the PD space and a space of modified correlation excitations.
We submit that in 2D electron systems with
correlations that are locally strong, even at the edge, a
tunneling electron can be accommodated, at low energies, only by
creating a periphery deformation.  We base this assertion on the 
observed absence of tunneling at low energies in the bulk of a strongly
correlated 2D system at any filling factor.
The PD state for a deformation
which accommodates charge $Q$ and is centered on the origin is
\begin{mathletters}
\beq\label{electron}
\ket{\psi[\eta]} = \exp{\Big\{ i\,\sum_{q>0} \left[ \eta_q\,\hat
\rho^{\text{1D}}_q +\bar\eta_q\,\hat\rho^{\text{1D}}_{-q}\right]
\Big\}}\, \ket{\Psi_{\text{0}}} \quad ,
\eeq
where
\beq
\lim_{q \to 0} \eta_q = -i\,\frac{2\pi}{\nu q L}\,\frac{Q}{e}
\quad .
\eeq
\end{mathletters}
We remark that $\ket{\psi[\eta]}$ will contain states outside of
the PD subspace if $\hat\rho^{\text{1D}}_{-q}$ for positive $q$
does not annihilate $\ket{\Psi_0}$, but these will not
contribute\cite{highenergy} to the zero-temperature low-energy
projected spectral weight of $\ket{\psi[\eta]}$. For a state
containing one added electron, the latter is readily calculated
by evaluating the inner products between $\ket{\psi[\eta]}$ and
PD-subspace Hamiltonian eigenstates. We find that 
\beq\label{final}
A^>(\varepsilon) \propto \sum_{M=1}^{\infty} \frac{\Gamma(M +
\nu^{-1})}{M! \,\Gamma (\nu^{-1})} \,\, \delta\Big(\varepsilon -
\frac{2\pi M}{L}\,\nu\,\hbar v_{\text{F}}\Big)\,\, ,
\eeq
where $\Gamma(\zeta)$ denotes Euler's Gamma function. It is easy
to see that Eq.~(\ref{final}) is the finite-size
equivalent\cite{jujo:prl:96} of Eq.~(\ref{powerlaw}).

Our proposal can be described in terms of a common cartoon for
the microscopic physics which underlies generalized $\chi$LL
models for quantum-Hall edge physics. The various $\chi$LL
branches can be associated with compressible strips in the
boundary layer between an incompressible bulk and vacuum. A
tunneling electron is assumed to be able to join any of these
branches. We are suggesting that, for the strongly correlated
edges which seem to be realized in recent experiments, adding an
electron to a particular incompressible strip would strongly
disrupt electronic correlations and, as occurs in the bulk at all
filling factors, yield states with no low-energy projection. On
the other hand, additional charge can be accommodated in periphery
deformations with little change in local correlation functions.  
If this is the case we are led, for $Q=e$, to identify 
the operator acting on $\ket{\Psi_0}$ in Eq.~(\ref{electron}) to
yield $\ket{\psi[\eta]}$ as the low-energy projection of the 
electron creation operator. This object is indeed reminiscent of
the bosonized electron creation operator in a conventional 1D
electron system.\cite{operator,voit:reprog:94} (A related
expression for the 2D electron operator in terms of density
fluctuations can be derived in a hydrodynamic
approach.\cite{glaz:prl:95})  We observe that this operator
does {\em not necessarily\/} obey the Fermi-statistics anticommutation
relation which is satisfied by the unprojected electron creation
operator. This contrasts with generalized $\chi$LL theories of
incompressible edges, where statistical compliance is used to
limit the objects which can appear in expressions for the
projected electron creation operator. If our suggestion is
correct, this requirement is not generic. Indeed, it is not
obvious that it should be since, at least for correlated edges, 
it is not possible to exchange electrons within the low-energy
sector of an edge-state theory. Also, the absence of a non-interacting
system that is adiabatically connected to a $\chi$LL obviates the
requirement to recover such a limit in bosonized expressions for
correlation functions. Our conclusion appears to be
similar to that reached in a recent fieldtheoretic
study.\cite{ammp} On the other hand, plausible arguments can be
advanced\cite{fisherleewen} that the temperature scales reached
by current experiments are still too high for the increase in
tunneling exponents forced by Fermi-statistics compliance to be
visible, although a rough estimate based on accurate
evaluations\cite{twothird} of edge-mode velocities for an abrupt
edge at $\nu = 2/3$ suggests that this cut-off temperature should
be $\sim 0.5$~K, which is an order of magnitude larger than the
experimental base temperature. Experiments on samples with
tunable confinement potentials could resolve this question more
decisively.

We thank A.~M.~Chang, J.~P.~Eisenstein, M.~P.~A.~Fisher,
M.~R.~Geller, S.~M.~Girvin, L.~I.~Glazman, and M.~Grayson for
illuminating discussions. This work was funded in part by NSF
Grant No.~DMR-9714055. U.Z.\ acknowledges support
from Studienstiftung des deutschen Volkes (Bonn, Germany).


\widetext
\end{document}

%% file: uzspecial.tex
\renewcommand{\narrowtext}{\begin{multicols}{2} \global\columnwidth20.5pc}
\renewcommand{\widetext}{\end{multicols} \global\columnwidth42.5pc}

\input{epsf.tex}

\def\top#1{\vskip #1\begin{picture}(290,80)(80,500)\thinlines \put(
65,500){\line( 1, 0){255}}\put(320,500){\line( 0, 1){
5}}\end{picture}}
\def\bottom#1{\vskip #1\begin{picture}(290,80)(80,500)\thinlines \put(
330,500){\line( 1, 0){255}}\put(330,500){\line( 0, -1){
5}}\end{picture}}

%% file: paper.bbl
\begin{thebibliography}{10}
\bibitem{spinapology}
For clarity, we neglect the electronic spin degree of freedom in
the discussion presented here.
\bibitem{circsym}
It is convenient to assume circular symmetry so that total
angular momentum is a good quantum number.
\bibitem{qell}
Angular momentum in the symmetric gauge corresponds to linear
momentum in $2 \pi / L$ units in the Landau gauge.  Thus the
assumption $M \ll \sqrt{N}$ corresponds to $q \ell \ll \sqrt{N}
\, 2 \pi \ell / L \sim 1$, assuming a roughly square Hall bar.
\bibitem{highenergy}
This is most easily seen in first quantization where
$\hat\rho^{\text{1D}}_{-q}$ for positive $q$ can be written
as a sum of derivatives with respect to complex coordinates.
These operators lower the relative angular momenta of pairs of
electrons and create high-energy states.
\end{thebibliography}
